\documentclass[a4paper,prd,preprintnumbers,twocolumn,superscriptaddress,nofootinbib,amsmath,amssymb]{revtex4-2}

\usepackage[dvips]{graphics}
\usepackage{color,hyperref}
\usepackage{times}
\usepackage{mathrsfs}
\hypersetup{colorlinks=true,linkcolor=blue,citecolor=red,filecolor=magenta,urlcolor=blue}

\usepackage{enumitem}
\usepackage{graphicx,subfigure}
\usepackage{dcolumn}
\usepackage{bm}
\usepackage{color}

\begin{document}
\title{Exploring a Novel Feature of Ellis Spacetime: Insights into Scalar Field Dynamics}
\author{Bobur~Turimov}
\email{bturimov@astrin.uz}
\affiliation{Institute of Fundamental and Applied Research, National Research University TIIAME, Kori Niyoziy 39, Tashkent 100000, Uzbekistan}
%\affiliation{Ulugh Beg Astronomical Institute, Astronomy St. 33, Tashkent 100052, Uzbekistan}

\author{Akbar Davlataliev}
\email{akbar@astrin.uz}
%\affiliation{Institute of Fundamental and Applied Research, National Research University TIIAME, Kori Niyoziy 39, Tashkent 100000, Uzbekistan}
\affiliation{National University of Uzbekistan, Tashkent 100174, Uzbekistan}

\author{Bobomurat Ahmedov}
\email{ahmedov@astrin.uz}
\affiliation{Institute of Fundamental and Applied Research, National Research University TIIAME, Kori Niyoziy 39, Tashkent 100000, Uzbekistan}
\affiliation{Department of Physics and Mathematics, Uzbekistan Academy of Sciences, Y. Gulomov 70, Tashkent 100047, Uzbekistan}

\author{Zden\v ek  Stuchl\'ik}
\email{zdenek.stuchlik@physics.slu.cz}
\affiliation{Research Centre for Theoretical Physics and Astrophysics, Institute of Physics, Silesian University in Opava, Bezrucovo n\'am. 13, CZ-74601 Opava, Czech Republic}

\date{\today}
\begin{abstract}
We have studied neutral and charged massive particles dynamics in Ellis spacetime in the presence of the external scalar field. Focusing on the circular motion of massive particles, the impact of an external scalar field on the Innermost Stable Circular Orbit (ISCO) position is analyzed, revealing a non-linear relationship with the scalar field parameter. Perturbation techniques are employed to investigate oscillatory motion near stable orbits in the Ellis spacetime, yielding analytical expressions for radial and angular oscillations. The throat of the wormhole has been constrained by comparing theoretical and observational results for fundamental frequencies of particles from quasars. Finally,  scalar and gravitational perturbations in the Ellis spacetime have been studied. It is shown that the equation for the scalar profile function is fully independent from the tensor functions and the solution can be represented in terms of the confluent Heun function. However, it has been shown that equations for the tensor profile functions strongly depend on the scalar profile functions in the Ellis spacetime and they are reduced to the Regge–Wheeler–Zerilli equation. Finally, numerical solutions to the Regge–Wheeler–Zerilli equation for the radial functions in the Ellis have been presented.

\end{abstract}

\maketitle

\section{Introduction}

Wormholes are hypothetical tunnels in spacetime connecting two separate regions of the universe or even different universes. They are often depicted in science fiction as shortcuts through space and time, allowing for rapid travel between distant points or even different dimensions. These hypothetical tunnels in space-time are solutions to the equations of general relativity proposed by Albert Einstein. However, while wormholes are theoretically possible according to the laws of physics, they remain purely speculative as no observational evidence of their existence has been found yet. Despite their speculative nature, wormholes have captured the imagination of scientists, appearing frequently in science fiction as portals for interstellar travel or time travel. They remain an intriguing topic of study in theoretical physics, pushing the boundaries of our understanding of space-time and the universe. 

The concept of a wormhole was initially introduced by John Wheeler in \cite{Wheeler1955PR,Misner1957AP}, who reinterpreted the Einstein-Rosen bridge as a link between distant points in spacetime without direct interaction \cite{Einstein1935PR}. In Ref. \cite{Morris1988AJP,Morris1988PRL} it has been proposed a basic metric for a wormhole that, theoretically, could be traversed by humans. Since then, numerous publications have explored various types of wormholes, but they all share the common trait of violating the weak energy condition. For a more thorough examination, refer to works such as \cite{}.

The Ellis wormhole, also known as the Bronnikov-Ellis wormhole \cite{Ellis1973JMP} or the Morris-Thorne wormhole \cite{Morris1988AJP}, refers to a theoretical construct in the field of general relativity. It's a type of wormhole solution derived from Einstein's field equations, which describe the gravitational interaction between matter and spacetime. The Bronnikov-Ellis wormhole offers a mathematical description of a hypothetical tunnel-like structure in spacetime. 

Another type of wormhole solution also known as the naked singularity has been studied in \cite{Janis1968PRL}. In Ref.\cite{Lobo2020PRD} the construction of the related spherically-symmetric thin-shell traversable wormholes within the framework of general relativity has been studied. A "defect wormhole" solution has been extensively discussed in\cite{Baines2023Univiverse}. General wormhole solutions within Einstein gravity, featuring an exponential shape function around both an ultrastatic and a finite redshift geometry have been studied in Refs.\cite{Papapetrou1954ZP,Dutta11102236D,Boonserm2018PRD,Turimov2022PDU}. Specific traversable wormhole solutions in Einstein-Dirac-Maxwell theory without including any exotic matter have been discussed in \cite{Salcedo2021PRL}. Charged wormhole solution in Einstein-Maxwell-scalar field theory has been studied in \cite{Turimov2021PDU}. Three parametric wormhole solutions in Einstein-scalar field theory have been investigated in \cite{Turimov2022particles}. In Ref. \cite{Chen-Hao240411002H}, the spherically symmetric traversable wormhole solutions in the presence of the scalar field supported by a phantom field in the AdS spacetime have been studied. Possible wormhole solutions in the Friedmann universe have been studied in \cite{Bronnikov2023Universe}. A magnetized dusty wormhole is discussed in \cite{Bronnikov2021Univ}. The exact geometric optics problems in the Ellis wormhole spacetime, namely, gravitational lensing effects including the computation of the deflection angle, have been studied in \cite{Muller2008PRD,Ghaffarnejad2022GReGr,Zhang2024PhRvD,Davlataliev2023pdu,Alloqulov2024ChPhC,Shaikh2023JCAP,Manna2023NewA}

In Ref. \cite{Deligianni2021PRD}, an analysis is conducted on the characteristics of circular orbits for massive particles situated in the equatorial plane of symmetric rotating Ellis wormholes. This includes the determination of orbital frequencies as well as radial and vertical epicyclic frequencies. Additionally, in Ref. \cite{Deligianni104b4048D,Vrba2023EPJC}, the study explores quasi-periodic oscillations emanating from the accretion disk surrounding rotating traversable wormholes. The focus lies on examining the linear stability of circular geodesic orbits in the equatorial plane across a broad class of wormhole geometries, leading to the derivation of analytical expressions for the epicyclic frequencies. The motion of spinning test particles around a wormhole is investigated in Ref.\cite{Benavides104h4024B} using the Mathisson-Papapetrous-Dixon (MPD) equations, which couple the Riemann tensor with the anti-symmetric tensor.

A slowly rotating Ellis-Bronnikov wormhole has been discussed in \cite{Azad240308387A}. The gravitational perturbations of the Morris-Thorne wormhole expanding up to the second order in rotation has been studied in Ref.\cite{Kang37j5012K}. It outlines the derivation process using the Newman-Penrose formalism and demonstrates the application of the Teukolsky equation to the wormhole spacetime. The article also details the computation of the perturbed Weyl scalars and the acquisition of its master equation. In Ref.\cite{Cremona101j4061C}, it has been introduced a comprehensive technique for deriving a gauge-invariant wave system of linearized perturbation equations in the context of the spherically symmetric Ellis-Bronnikov wormhole.

{In general relativity, the motion of particles in curved spacetime is a fundamental concept, elucidating how matter and energy influence the curvature of spacetime and, consequently, how this curvature governs the trajectories of objects. In such a curved spacetime, particles traverse paths known as geodesics, which are determined by the geodesic equation. However, when a charged particle is subjected to an external electromagnetic field, it experiences acceleration and emits electromagnetic radiation, a phenomenon described by a non-geodesic equation. Various forms of electromagnetic radiation and their corresponding mathematical formulations are detailed in \cite{Landau-Lifshitz2}. Furthermore, the study of particle dynamics in the presence of an external scalar field is both intriguing and significant. The interaction between massive particles and scalar fields remains a topic of uncertainty, yet it is theoretically vital for comprehending astrophysical processes near compact objects such as black holes and neutron stars. This issue has been examined in Ref. \cite{Noble2021NJP}, which includes the analysis of self-force effects in the presence of an external scalar field. Certain facets of this self-force can be interpreted through the geometry of spacetime, offering insights into the paradox of a particle radiating without experiencing a self-force. The acceleration of particles by black holes in the presence of a scalar field has been investigated in \cite{Zaslavskii2017IJMPD}. The interaction between scalar field and massive particle is also introduced by Misner et. al.~\cite{Misner1972PRL} and by Breuer at. al. \cite{Breuer1973PRD} to describe scalar perturbation or so-called geodesic synchrotron radiation in the Schwarzschild spacetime.}

In the present paper, we are interested in testing the Ellis spacetime by considering particle motion around the wormhole in the presence of the external scalar field, using the technique mentioned in Ref. \cite{Noble2021NJP}. The paper is organized as follows: In Sec.~\ref{Sec2}, we provide the main equations that are related to background spacetime and dynamics motion of test particle in the presence of the external scalar field. In Sec.~\ref{Sec3}, we study particle motion including radiation reaction. In Sec. \ref{Sec4}, we study wormhole perturbation. Finally, in Sec.~\ref{Sec:conclusions} we summarize obtained results. Throughout the paper, we use the geometrized system of units in which $c=G=\hbar=1$ and spacelike signature ($-,+,+,+$).
%%%

\section{Background geometry and particle dynamics}\label{Sec2}

\begin{figure}
\centering\includegraphics[width=\hsize]{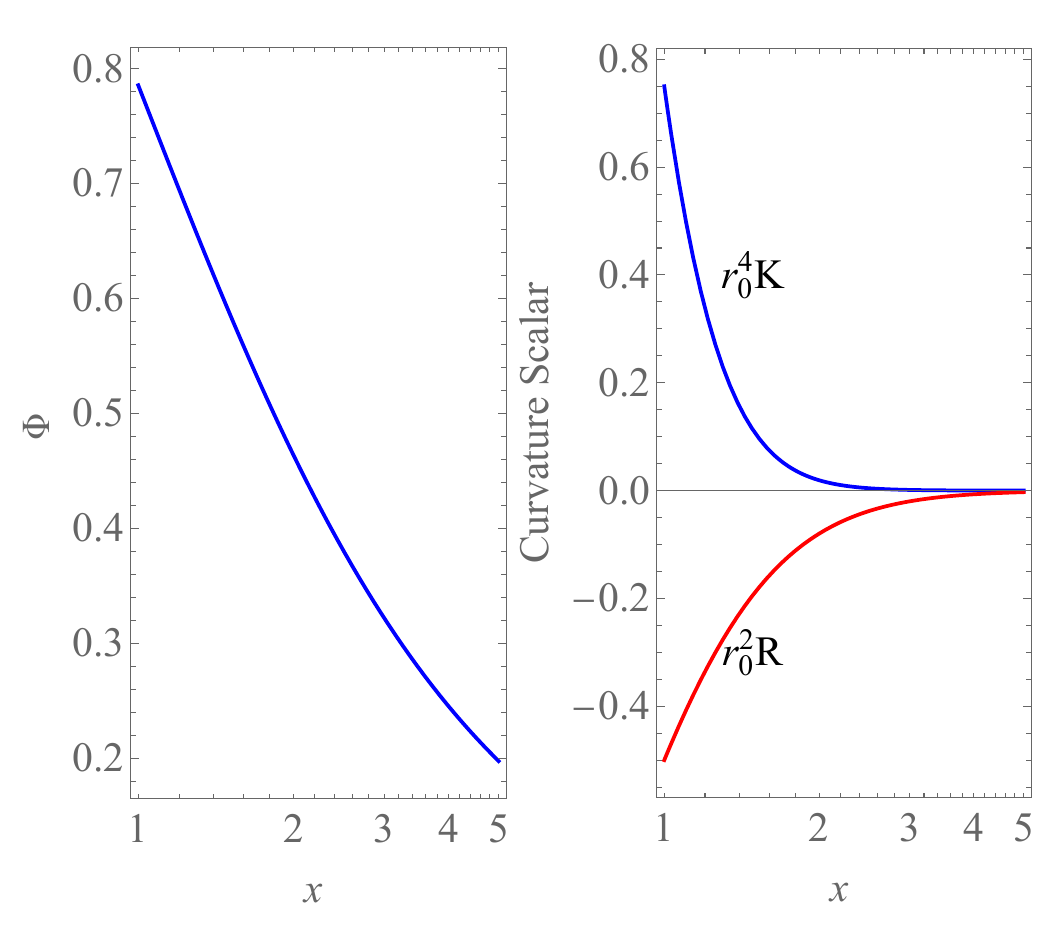}
\caption{Radial dependence of the scalar field $\Phi(x)$ and curvature invariant, dimensionless Ricci scalar $r_0^2R$ and dimensionless Kretschmann scalar $r_0^4K$.}
    \label{Scalar}
\end{figure}

The Ellis wormhole is governed by the spacetime line element \cite{Ellis1973JMP,Morris1988AJP}:
\begin{align}\label{metric}
&ds^2=-dt^2+dr^2+(r^2+r_0^2)(d\theta^2+\sin^2\theta d\phi^2)\ ,
\end{align}
along with the associated scalar field
\begin{align}
&\Phi=\frac{\pi}{2}-\tan^{-1}\left(\frac{r}{r_0}\right)\ ,\label{Phi}
\end{align}
where $r_0$ is the throat of the wormhole. Here the radial coordinate $r$ runs between $r_0$ and infinity, i.e. $r_0< r<\infty$. One has to emphasise that it is one of the simple wormhole solutions of the Einstein-scalar field equation with the following energy-momentum tensor: 
\begin{align}
&T^\mu_{\,\,\,\nu}=\frac{r_0^2}{\left(r^2+r_0^2\right)^2}\left(
\begin{array}{cccc}
 1 & 0  & 0 & 0 \\
 0 & -1 & 0 & 0 \\
 0 & 0  & 1 & 0 \\
 0 & 0  & 0 & 1 \\
\end{array}
\right) \ ,
\end{align}
while the curvature scalar invariants of the spacetime such as Ricci scalar and Kretschmann scalar can be expressed as
\begin{align}
&R=-\frac{2r_0^2}{(r^2+r_0^2)^2}\ ,\qquad K=\frac{12r_0^4}{(r+r_0^2)^4}\ ,    
\end{align}
which are regular at any point of spacetime at $r>r_0$. The radial dependence of the scalar field and curvature invariants, Ricci and Kretchmann scalar is shown in Fig.\ref{Scalar}.
\begin{figure}
\centering\includegraphics[width=0.49\linewidth]{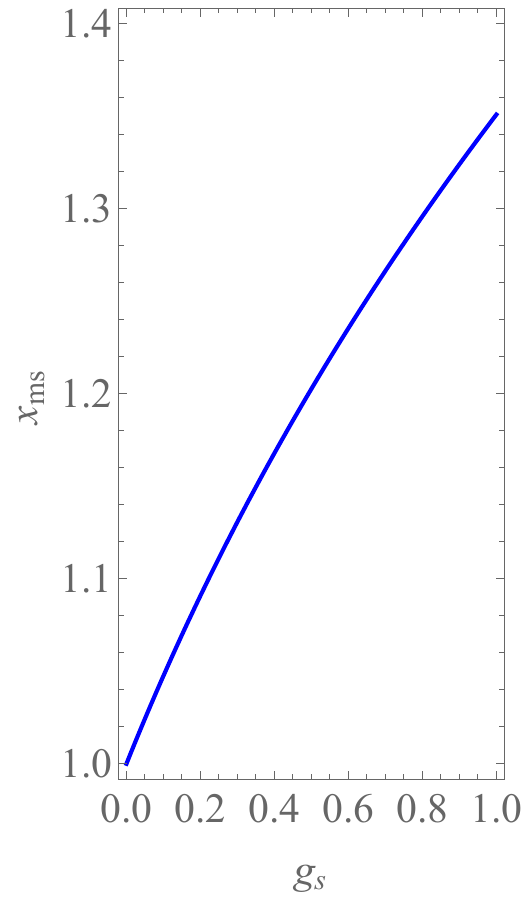}
\includegraphics[width=0.435\linewidth]{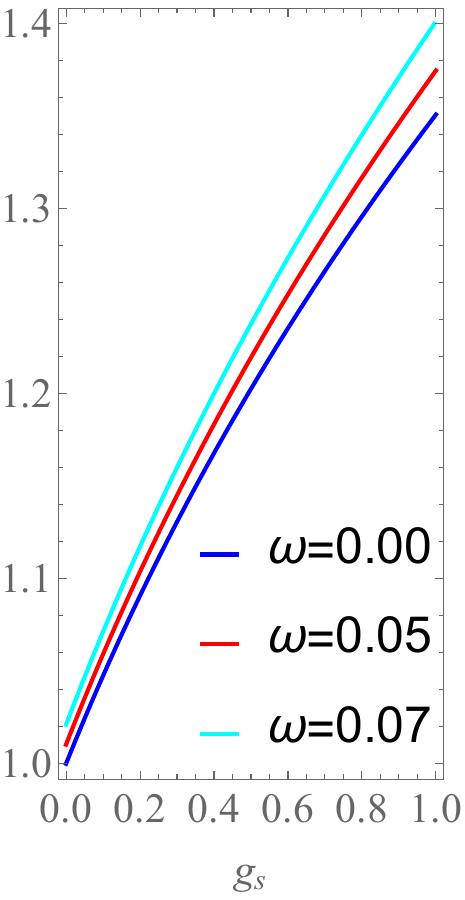}
\caption{(Left panel) Dependence of the marginally bound orbit circular orbit of neutral particle $x_{\rm ms}$ from the coupling constant $\text{g}_s$. (Right panel) Dependence of the marginally bound orbit circular orbit of charged particle $x_{\rm ms}$ from the coupling constant $\text{g}_s$ for the different values of the magnetic parameter.}\label{fig:ISCO}
\end{figure}
\begin{figure}
\centering\includegraphics[width=\hsize]{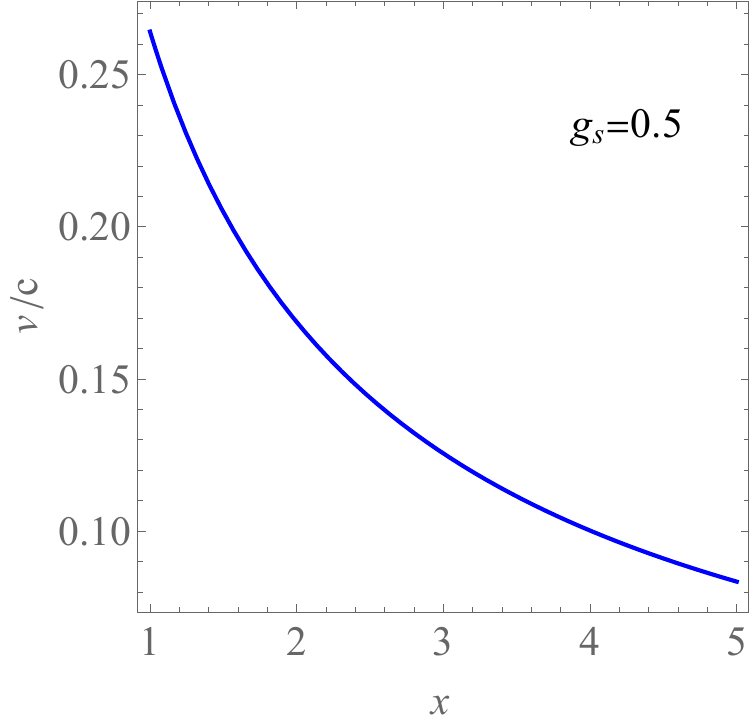}
\caption{Radial dependence of the linear orbital velocity .}\label{velocity}
\end{figure}
\begin{figure*}
    \centering
    \includegraphics[width=0.3\linewidth]{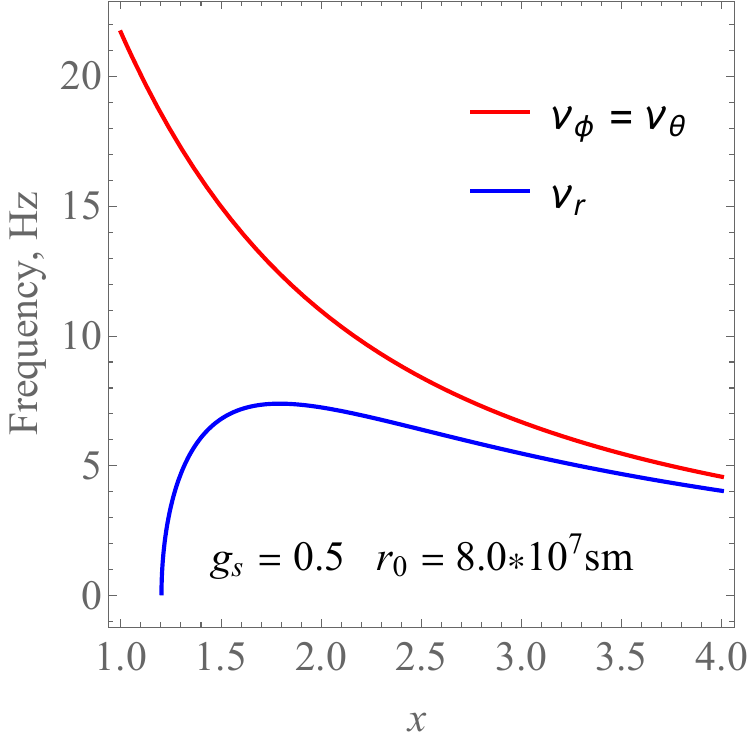}
    \includegraphics[width=0.31\linewidth]{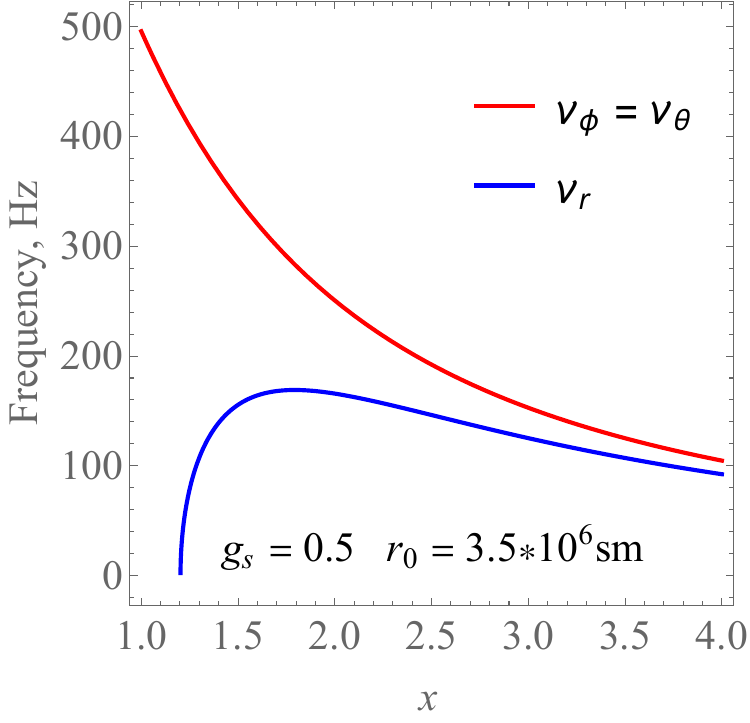}
    \includegraphics[width=0.32\linewidth]{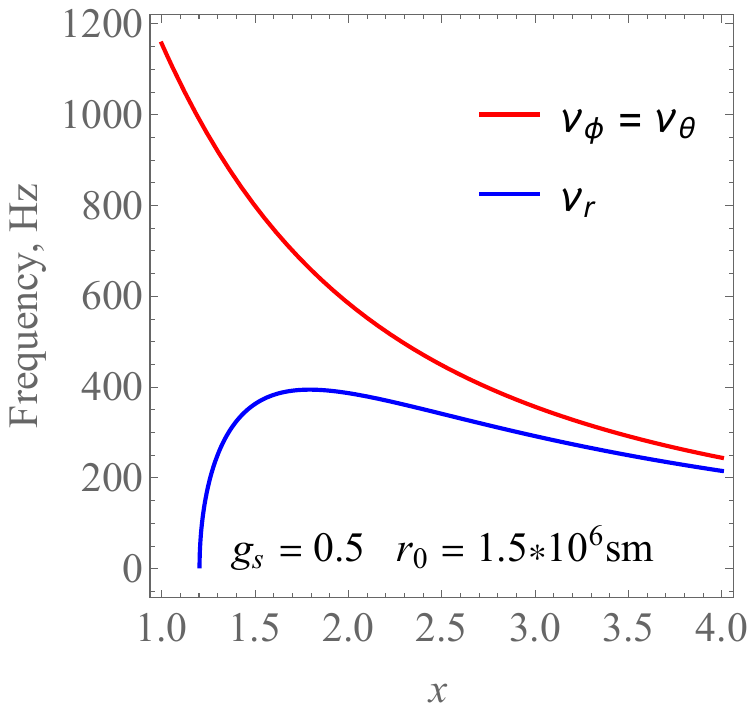}
    \caption{Radial dependence of the fundamental frequencies of a massive particle.}
    \label{nu}
\end{figure*}

We can immediately observe that the motion of the test particle in the Ellis spacetime background lacks interest due to the absence of temporal and radial components in the metric tensor. However, when incorporating the interaction between a test particle and an external scalar field, the problem becomes intriguing. In this case, the Lagrangian describing a test particle with mass $m$ in the presence of the external scalar field $\Phi$ can be expressed as~\cite{Noble2021NJP,Misner1972PRL,Breuer1973PRD,Turimov2024arXiv}
\begin{align}\label{Lag}
L=\frac{1}{2}m_*g_{\mu\nu}u^\mu u^\nu\ ,\qquad m_*=m(1+\text{g}_s\Phi)\ , 
\end{align}
where $m_*$ is the effective mass of a test particle in the presence of the external scalar field and influenced by a coupling constant, $\text{g}_s$, and $u^\mu=dx^\mu/d\tau$ is the four-velocity of test particle normalized as $u_\mu u^\mu=-1$. Here, $\tau$ is the particle's proper time. The constants of the motion namely, the specific energy ${\cal E}$ and specific angular momentum ${\cal L}$ of test particle are expressed as follows
\begin{align}\label{c}
{\cal E}=(1+\text{g}_s\Phi){\dot t}\ ,\qquad {\cal L}=(1+\text{g}_s\Phi)(r^2+r_0^2){\dot\phi}\ . 
\end{align}

Hereafter using normalization of the four-velocity of massive test particle (i.e. $u_\mu u^\mu=-1$) along constants of motion in the expression \eqref{c}, one can obtain
\begin{align}\label{V}
g_{rr}{\dot r}^2+g_{\theta\theta}{\dot\theta}^2+V(r,\theta)=0\ .
\end{align}
where $V(r,\theta)$ is defined as
\begin{align}
V(r,\theta)=1-\frac{1}{(1+\text{g}_s\Phi)^2}\left[{\cal E}^2-\frac{{\cal L}^2}{(r^2+r_0^2)\sin^2\theta}\right]\ ,
\end{align}
which limits constants of motion.

\subsection{Marginally stable circular orbit}

The concept of stable and unstable orbits of test particle in the vicinity of compact and massive objects such as black holes, and neutron stars are well-described. It is also interesting to consider the marginally bound and marginally stable circular orbit of test particle motion near the wormhole. For simplicity, one can consider circular motion in the equatorial plane (i.e. $\theta=\pi/2$ and ${\dot\theta}=0$), equation of motion reduces to
\begin{align}
&{\dot t}=\frac{{\cal E}}{(1+\text{g}_s\Phi)}\ ,
\\
&{\dot\phi}=\frac{\cal L}{(1+\text{g}_s\Phi)(r^2+r_0^2)}\ ,
\\
&{\dot r}^2 =\frac{1}{(1+\text{g}_s\Phi)^2}\left({\cal E}^2-\frac{{\cal L}^2}{r^2+r_0^2}\right)-1\ .
\end{align}
Using conditions ${\dot r}={\ddot r}=0$ and eliminating the specific angular momentum from the final expression, the critical value of the specific energy of a massive particle can be found as  
\begin{align}
&{\cal E}^2=\left(1+\text{g}_s\Phi -\text{g}_s\frac{r_0}{r}\right)\left(1+\text{g}_s\Phi\right)\ ,
\end{align}
which is equal to $1$ at the infinity. From this expression one can obtain two characteristic radii of particle, one is called marginally bound circular orbit and another one is marginally stable circular orbit.  

The marginally bound circular orbit of particle can be found from the condition ${\cal E}=1$ which does not exist around Ellis wormhole, while the stationary point of the specific energy is located at the marginally bound orbit of the massive particle (i.e. ${\cal E}'(r)=0$) and determined from the following equation:
\begin{align}\label{ISCO}
\text{g}_s x+\left(1-x^2\right)\left[1+\frac{\pi}{2}\text{g}_s-\text{g}_s\tan^{-1}(x)\right]=0\ ,
\end{align}
where $x=r/r_0>1$ is the dimensional radial coordinate. It is impossible to obtain the analytical solution to equation \eqref{ISCO}, however, but employing numerical methods can yield the position of the marginally stable circular orbit of massive particle around the wormhole. Our numerical investigations showed that such orbits are smaller in size compared to the wormhole's throat (i.e. $x<1$) when the coupling constant $\text{g}_s$ takes a negative value, which is meaningless. Conversely, as the coupling constant $\text{g}_s$ becomes positive, these orbits get larger (i.e. $x>1$). Figure \ref{fig:ISCO} illustrates how the marginally bound circular orbit of a test particle varies with the coupling constant.

\subsection{Fundamental frequencies}

Here we will discuss fundamental frequencies, namely, orbital and epicyclic frequencies, of massive particle in the Ellis spacetime. From the Lagrangian \eqref{Lag} equation of motion for a massive particle in the presence of the external scalar field yields \cite{Noble2021NJP}
\begin{align}\label{eom}
\frac{Du^\mu}{d\tau}=(g^{\mu\nu}+u^\mu u^\nu)\partial_\nu\ln\frac{m_*}{m}\ ,\qquad u_\mu\frac{Du^\mu}{d\tau}=0\ .   
\end{align}
Now considering the circular motion with the four-velocity of $u^\mu={\dot t}(1,0,0,\Omega)$ and angular velocity of test particle is derived from equation \eqref{eom} as follows: 
\begin{align}
\Omega=\frac{d\phi}{dt}=\frac{1}{r_0}\frac{1}{\sqrt{(x^2+1)\left[1+\frac{x}{\text{g}_s}\left(1+\text{g}_s\Phi\right)\right]}}\ ,    
\end{align}
while the linear orbital velocity of particle measured by a local observer yields
\begin{align}
v_\phi=\Omega\sqrt{\frac{g_{\phi\phi}}{-g_{tt}}}=\sqrt{\frac{1}{1+\frac{x}{\text{g}_s}\left(1+\text{g}_s\Phi\right)}}<1\ ,    
\end{align}
which is less than the speed of light for the positive value of the interaction parameter. The radial dependence of the linear velocity of massive particle orbiting around the Ellis wormhole in the presence of the scalar field is illustrated in Fig.\ref{velocity}. 

It is also an interesting and important task to consider the radial and vertical oscillatory motion of test particle around the stationary stable orbit. Using equation \eqref{V}, the radial and vertical epicyclic frequencies are \cite{Turimov2022Univ}   
\begin{align}
\Omega_r=\sqrt{\frac{1}{2g_{rr}{\dot t}^2}\frac{\partial^2V}{\partial r^2}}\ ,
\qquad 
\Omega_\theta=\sqrt{\frac{1}{2g_{\theta\theta}{\dot t}^2}\frac{\partial^2V}{\partial\theta^2}}\  ,
\end{align}
which satisfies the following $2D$-oscillator equations for displacements $(d^2/dt^2+\Omega_r^2)\delta r=0$ and $(d^2/dt^2+\Omega_\theta^2)\delta\theta=0$. To have an idea about the value of the fundamental frequencies of massive particle in the Ellis spacetime, we restore the fundamental constant in particular the speed of light. Our analyses showed that the vertical and orbital frequencies are equal to each other $\Omega=\Omega_\theta$. Now using the following notating $\nu_i=\Omega_i/2\pi$, measurable fundamental frequencies of massive particle can be determined as 
\begin{align}\label{nur}
\nu_r&=\frac{c}{2\pi r_0}\sqrt{\frac{(1+\text{g}_s\Phi)\left(x^2-1\right)-\text{g}_s}{\left(x^2+1\right)\left[1+\frac{x}{\text{g}_s}\left(1+\text{g}_s\Phi\right)\right]}}\ ,
\\\label{nutf}
\nu_\theta&=\nu_\phi=\frac{c}{2\pi r_0}\frac{1}{\sqrt{(x^2+1)\left[1+\frac{x}{\text{g}_s}\left(1+\text{g}_s\Phi\right)\right]}} \ .
\end{align}
As one can easily check that the frequencies are sensitive to the wormhole's throat. To compare the theoretical results with observational data we have restored the speed of the light in \eqref{nur} and \eqref{nutf}. As an astrophysical consequence, we make constraints on the throat of the Ellis wormhole by comparing the above expression with observational evidence of the fundamental frequencies. We can make constraints on wormhole parameters using observation evidence of the fundamental frequency for quasars and micro-quasars. In Fig.\ref{nu} radial dependence of the fundamental frequencies, namely, orbital, radial and vertical frequencies are illustrated.

\section{Charged particle dynamics}\label{Sec3}

Now we discuss the orbital motion of a charged particle around Ellis wormhole in the presence external uniform magnetic field. Assuming the wormhole is embedded in the asymptotically uniform magnetic field. By utilizing the Wald method \cite{Wald1974PRD}, it is possible to identify the elements of the four-vector potential that align with the magnetic field, in the following manner: $A^{\mu}=(0,0,0,B/2)$, where $B$ is the strength of the asymptotically uniform magnetic field. Consequently, the azimuthal component of the vector potential of the electromagnetic field is given by $A_\phi=(B/2)(r^2+r_0^2)\sin^2\theta$ and fully satisfies Maxwell equation. The Lagrangian for a charged particle in the presence of the external magnetic field is
\begin{align}
L=\frac{1}{2}m_*g_{\mu\nu}u^\mu u^\nu+q A_{\mu}u^{\mu} \ ,
\end{align}
where $q$ is a charge of the test particle. The Lagrangian equation of motion is derived as
\begin{align}
\frac{Du^\mu}{d\tau}&=\frac{q}{m}F^\mu_{\,\,\,\nu}u^\nu+(g^{\mu\nu}+u^\mu u^\nu)\partial_\nu\ln\frac{m_*}{m}\ ,   
\end{align}
and after simple algebraic manipulations equation of the radial motion yields
\begin{align}
{\dot r}^2 =\frac{1}{(1+\text{g}_s\Phi)^2}\left[{\cal E}^2-\left(r^2+r_0^2\right)\left(\frac{{\cal L}}{r^2+r_0^2}-\omega\right)^2\right]-1\ ,
\end{align}
where $\omega=qB/2m$ is a magnetic parameter. The marginally bound circular radius for a charged test particle takes using conditions $\dot{r}=\ddot{r}=0$:   
\begin{align}
\text{g}_s x+\left(1-x^2\right)\left(1+\text{g}_s\Phi\right)+4\omega^2 x^3(1+x^2)=0 \ .
\end{align}
Using the numerical calculation one can find the dependence of the marginally bound circular orbit radius of a charged particle orbiting wormhole from the interaction parameter $\text{g}_s$ from the different values of the magnetic parameter $\omega$ is shown in Fig.\ref{fig:ISCO}.  

\begin{figure*}
    \centering
    \includegraphics[width=0.32\linewidth]{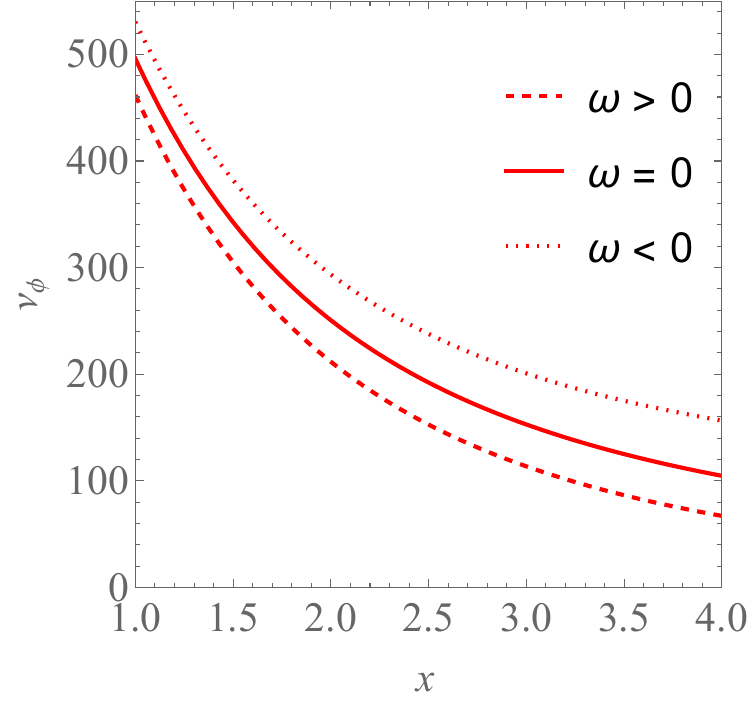}
    \includegraphics[width=0.32\linewidth]{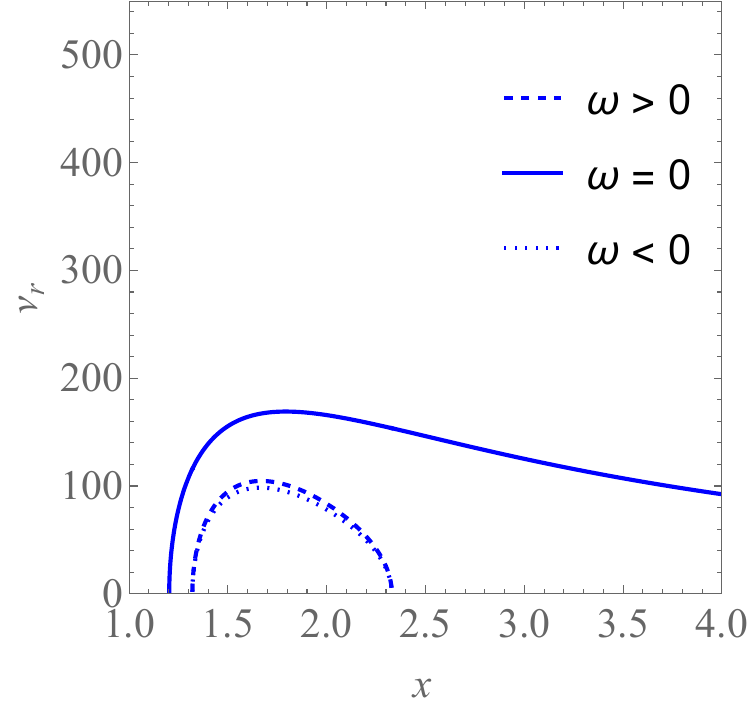}
    \includegraphics[width=0.32\linewidth]{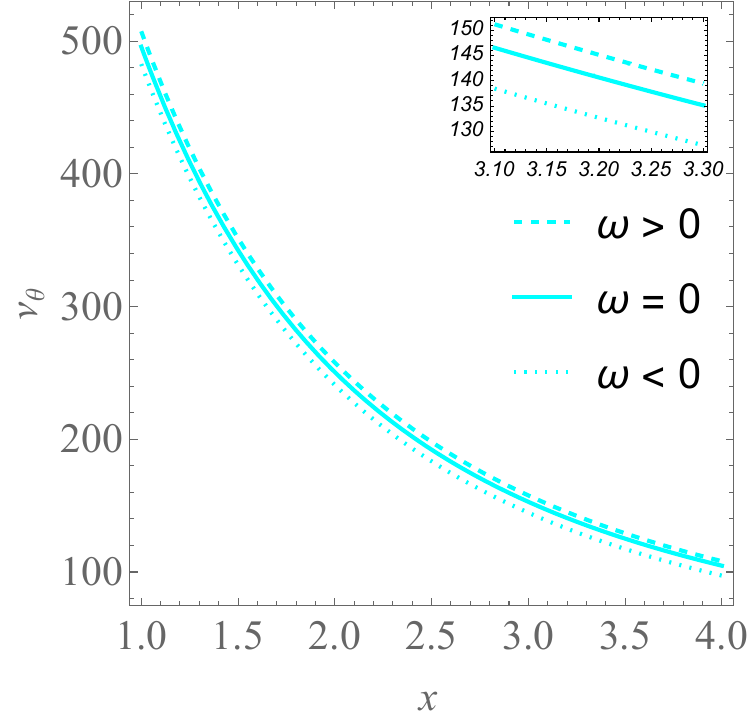}
    \caption{Fundamental frequencies of  charged particle around Ellis wormhole on  $\omega=\pm0.4$, parameter $\text{g}_s = 0.5$\ .}
    \label{nue}
\end{figure*}

The fundamental frequencies of the charged particle in the vicinity of Ellis wormhole in the presence of the external magnetic field can be easily done. The Keplerian frequency of charged particle is derived from the following expression:
\begin{align}
\Omega^2+\omega\Omega\sqrt{1-\Omega ^2(x^2+1)}=\frac{\text{g}_s\left[1-\Omega^2(x^2+1)\right]}{x(x^2+1)(1+\text{g}_s\Phi)}\ .
\end{align}
Similarly, epicyclic frequencies for charged particle orbiting around the Ellis wormhole in the presence magnetic field, however we will skip the detailed calculations. The radial dependence of the fundamental frequencies is shown in Fig. \ref{nue}. As one can see from this figure the fundamental frequencies of charged particle are split due to the external magnetic field. On the other hand, our results show  that Keplerian frequencies strongly depend on the magnetic parameter. 

Here we will discuss about radiation from massive particle in the Ellis spacetime in the presence of the scalar field. From the equation of motion \eqref{eom} one can see that the four-acceleration is non-zero $w^\mu=Du^\mu/d\tau$. The intensity of the radiating particle is proportional to the square of the acceleration of particle:
\begin{align}
I\sim\frac{Du_\mu}{d\tau}\frac{Du^\mu}{d\tau}&=\frac{q}{m}F_{\mu\lambda}F^\mu_{\,\,\,\nu}u^\nu u^\lambda+(g^{\mu\nu}+u^\mu u^\nu)\partial_\mu\frac{m_*}{m}\partial_\nu \frac{m_*}{m}\ .      
\end{align}
Here the first term of the last equation denotes electromagnetic radiation from charged particle while the second term is the intensity of radiation from massive particle due to interaction with the external scalar field.

Now we consider charged particle motion including the radiation reaction along massive particle trajectories in the presence of the scalar field. The Lorentz-Abraham-Dirac equation can be expressed as
\begin{align}\label{RadEq}
\frac{Du^\mu}{d\tau}&=\frac{q}{m}F^\mu_{\,\,\,\nu} u^\nu+(g^{\mu\nu}+u^\mu u^\nu)\partial_\nu\ln \frac{m_*}{m}+\frac{1}{2}\tau_0\left(R^\mu_{~\nu}+u^\mu u_\lambda R^\lambda_{~\nu}\right)u^\nu+\tau_0\ ,   
\end{align}
where $R_{\mu\nu}$ represents the Ricci tensor. Parameter $\tau_0$ is rather small parameter than all other terms therefore it can serve as an expansion parameter. Hereafter applying the Landau trick equation of motion in the presence of the external scalar field reads 
\begin{align}\label{LandauEq}\nonumber
\frac{Du^\mu}{d\tau}&=\frac{q}{m}F^\mu_{\,\,\,\nu} u^\nu+h^{\mu\nu}\partial_\nu\ln\frac{m_*}{m}\\&+\tau_0\left[\frac{1}{2}h^\mu_\lambda R^\lambda_{~\nu} u^\nu+h^{\mu\nu}\left(\frac{D}{d\tau}+u^{\alpha}\partial_{\alpha}\ln\frac{m_*}{m}\right) \partial_{\nu}\ln\frac{m_*}{m}+\frac{q}{m}\left(u^\alpha\nabla_\alpha F_{~\nu}^\mu+\frac{q}{m}h^{\mu\lambda}F_{\lambda\alpha}F^\alpha_{~\nu}\right)u^\nu\right]\ ,  
\end{align}
where $h^{\mu\nu}=g^{\mu\nu}+u^\mu u^\nu$. 

\begin{figure}
    \centering
    \includegraphics[width=\hsize]{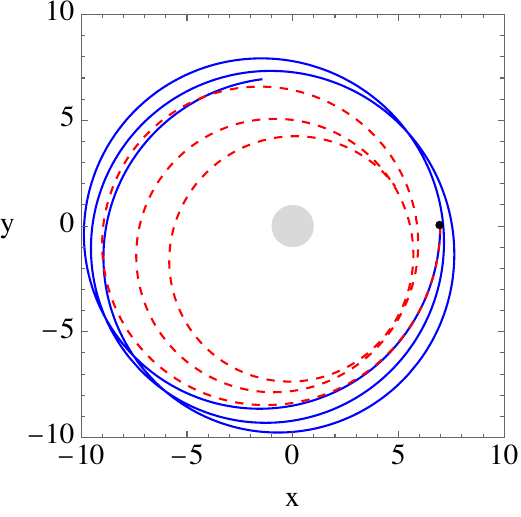}
    \caption{Particle trajectories near Ellis wormhole  the radiation trajectory without ($\tau_0=0$, solid blue line) and with ($\tau_0=0.05$, dashed red line) radiation reaction term. $\text{g}_s=0.4$ and $\omega=0.1$}
    \label{radiation}
\end{figure}

The fundamental concept behind transforming equation \eqref{RadEq} into the form depicted in \eqref{LandauEq} is to represent it as a third-order system of differential equations concerning four coordinates. This implies the necessity of determining twelve constants of motion, a pivotal aspect in understanding particle dynamics within curved spacetime. However, avoiding this issue is achievable through the Landau trick, resulting in the equation of motion being simplified to a second-order system concerning the four coordinates $x^\mu$. While the explicit expressions for each coordinate in \eqref{LandauEq} are considerably lengthy, they will not be elaborated upon in this correspondence. Nevertheless, through careful numerical analysis, employing prescribed initial conditions: $(0,r_i,\theta_i,0)$ and $\left\{-{\cal E},0,0,{\cal L}/(r_i^2+r_0^2)\sin^2\theta_i\right\}$, it becomes feasible to ascertain the dependency of each coordinate on the affine parameter, denoted as $x^\mu=x^\mu(s)$, where $(r_i, \theta_i)$ are the initial position of test particle. Subsequently, by implementing a coordinate transformation such as $x=r\cos\phi$a and $y=r\sin\phi$, the trajectories of particles can be visualized in Cartesian coordinates through parametric plots. The trajectory of particle in the Ellis spacetime in the presence of the scalar field is shown in Fig.\ref{radiation}. Interestingly, it turns out that radiation allows massive particle to escape the wormhole by exploiting the absence of a gravitational field. However, in Ref.\cite{Turimov2023PDU} it is shown that due to radiation particle starts to fall onto the black hole in the scalar-tensor-vector gravity.

\section{Wormhole perturbation}\label{Sec4}

The black hole perturbation is one of the hot topic, in particular, after LIGO and Virgo collaboration start detect the gravitational wave from the binary systems. This kind of perturbation cab also be applicable to wormhole spacetime. Here we are interested in considering perturbation of the Ellis spacetime. We have mentioned before that the Ellis spacetime is the solution of Einstein-scalar field equations. So that the scalar field and metric tensor can be expressed as $\tilde{\Phi}=\Phi+\delta\Phi$ and $\tilde{g}_{\mu\nu}=g_{\mu\nu}+h_{\mu\nu}$, where $\Phi$ and $g_{\mu\nu}$ are the scalar field and background spacetime metric given in Eqs. \eqref{metric} and \eqref{Phi}, while $\delta\Phi$ is perturbed scalar field is
\begin{align}
&\delta\Phi=e^{-i\omega t} F(r)P_\ell(\cos\theta)\ ,  
\end{align}
and perturbed metric tensor components  are defined as~\cite{Regge1957PR}
\begin{align}\label{gper}
&h_{\mu\nu}=e^{-i\omega t}\left(
\begin{array}{cccc}
 H_0 & H_1 & 0               & 0 \\
 H_1 & H_2 & 0               & 0 \\
 0      & 0      & K(r^2+r_0^2) & 0 \\
 0      & 0      & 0               & K(r^2+r_0^2)\sin^2\theta \\
\end{array}
\right)P_\ell(\cos\theta)  \ ,
\end{align}
where $H_0(r)$, $H_1(r)$, $H_2(r)$, $K(r)$ and $F(r)$ are unknown radial functions and $P_\ell=P_\ell(\cos\theta)$ is the Legendre polynomial which satisfies the following equation:
\begin{align}\label{P}
\frac{1}{\sin\theta}\frac{d}{d\theta}(\sin\theta\frac{d}{d\theta})P_\ell+\ell(\ell+1)P_\ell=0\ .    
\end{align}
After applying the property of the Legendre polynomial \eqref{P}, the explicit field equations for the radial functions are presented in Appendix \ref{App}. Notice that referring to Ref.\cite{Regge1957PR}, it's observed that the perturbation described in equation \eqref{gper} is termed as the "polar perturbation" while another perturbation, known as the "axial perturbation," is mentioned for the metric tensor in the reference. However, our investigation confirms the absence of this axial perturbation in the Ellis spacetime.  

\subsection{Time-independent solutions}

We first focus on finding the stationary solutions for the radial profile functions (i.e. $\omega=0$). In this case, the function $H_1$ vanishes, (i.e. $H_1=0$), and from Eqs. \eqref{eq1}-\eqref{KG}, one can obtain equations for the remaining functions in the form: 
\begin{align}\label{H}
&[\left(x^2+1\right) H']'-\ell(\ell+1)H=0\ ,
\\\label{F}
&[\left(x^2+1\right)F']'-\ell(\ell+1)F+\frac{4}{x^2+1}F=0\ ,
\\\label{K}
&K=H-\frac{4}{(\ell+2)(\ell-1)}\left(F'+\frac{2x}{x^2+1}F\right)\ ,
\end{align}
where primes denote the derivative with respect to the dimensionless radial coordinate $x$. Solutions to Eqs. \eqref{H} and \eqref{F} are rather simple and can be expressed as read
\begin{align}\label{Heq}
&H(x)=i^\ell\left[C_{1\ell}P_\ell(ix)+C_{2\ell}iQ_\ell(ix)\right]\ ,
\\\label{Feq}
&F(x)=i^\ell\left[C_{3\ell}P_{\ell}^2(ix)+C_{4\ell} iQ_{\ell}^2(ix)\right]\ ,    
\end{align}
where $C_{1\ell}$, $C_{2\ell}$, $C_{3\ell}$ and $C_{4\ell}$ are constants of integration. $P_\ell(ix)$, $Q_\ell(ix)$ are Legendre functions of the first, and second kinds while $P_\ell^2(ix)$ and $Q_\ell^2(ix)$ are associated Legendre functions of the first and second kind, respectively. Note that the coefficients $i^\ell$ in equations \eqref{H} and \eqref{F} ensure that the solution remains real at all points in spacetime. Finally, from Eq.\eqref{K} the function $K$ can be found as

\begin{align}
K(x)&=C_{1\ell}P_\ell\left(ix\right)+C_{2\ell}Q_\ell\left(ix\right)+\frac{4C_{3\ell}}{(\ell+2)\left(x^2+1\right)}[xP_\ell^2(ix)+iP_{\ell+1}^2(ix)]+\frac{4C_{\ell}4}{(\ell+2)\left(x^2+1\right)}[xQ_\ell^2(ix)+iQ_{\ell+1}^2(ix)]\ .   
\end{align}

\subsection{Wave solution}

\begin{figure}
\centering\includegraphics[width=\hsize]{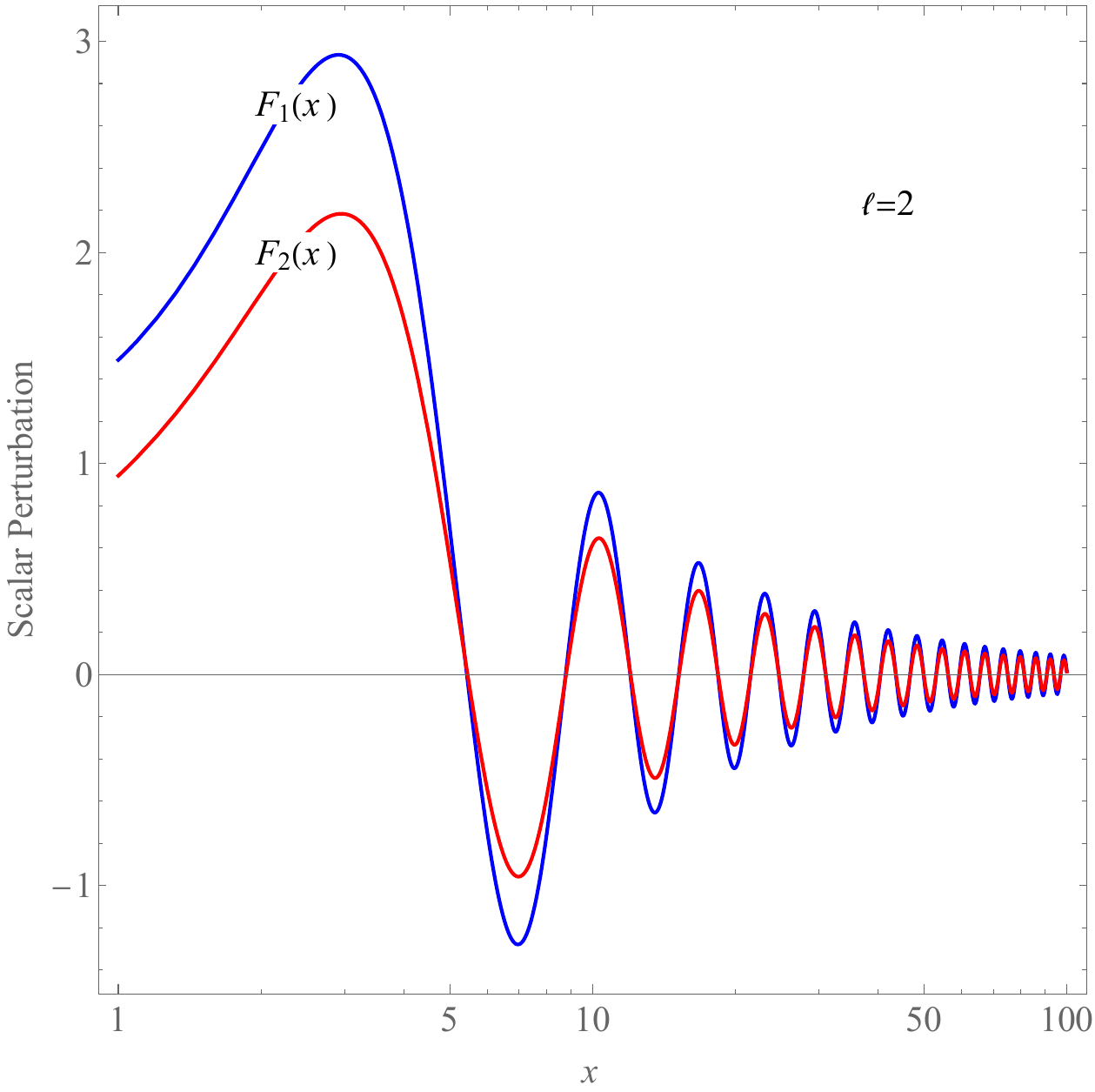}
\caption{Radial dependence of the scalar perturbation at $\ell=2$.\label{figF}}
\end{figure}

Here we will discuss the wave solution of the field equations for radial functions. Before going further we introduce a dimensionless frequency which is defined as $\omega_0=\omega r_0$. Now using expressions \eqref{KG1}, \eqref{KG} and taking into account the fact that $H=H_0=H_2$, the equation for the function $F$ can be easily derived as
\begin{align}\nonumber
&\left[\left(x^2+1\right)F'\right]'+\left[\omega_0^2x^2+\frac{4}{x^2+1}-4\eta\right]F=0\ ,
\end{align}
where $\eta=\ell^2+\ell-\omega_0^2$ and the analytical solution to the above equation can be presented in terms of the confluent Heun function (i.e. ${\rm HeunC}(a,b,c,d,x)$) as follows:
\begin{align}\label{Fpert}
&F(r)=\left(x^2+1\right)\left[c_1F_{1\ell}(x)+c_2xF_{2\ell}(x)\right]\ ,
\end{align}
where 
\begin{align}
F_{1\ell}(x)=\text{HeunC}\left[\eta-\frac{3}{2},-\frac{\omega_0^2}{4},\frac{1}{2},3,0,-x^2\right]\ ,
\\
F_{2\ell}(x)=\text{HeunC}\left[\eta-3,-\frac{\omega_0^2}{4},\frac{3}{2},3,0,-x^2\right]\ .
\end{align}
As we found that in the Ellis spacetime, the solution to scalar perturbation is described by the analytical expression, namely, confluent Heun function. The radial dependence of the function $F(x)$ is shown in Fig. \ref{figF}. The radial profile function $F(x)$ oscillates while simultaneously decreasing. 

Now we concentrate on the equations for the remaining radial functions $H(r)$, $H_1(r)$ and $K(r)$. The components of Einstein field equations for those functions yield
\begin{align}
K''&+\frac{3x}{x^2+1}K'-\frac{\ell^2+\ell-2}{2\left(x^2+1\right)}K-\frac{x}{x^2+1}H'-\frac{\ell^2+\ell+2}{2\left(x^2+1\right)}H=-\frac{2}{x^2+1}F'\ ,\\
K'&-H'-\frac{\ell(\ell+1)-2}{2x}(K-H)+\frac{\omega_0^2\left(x^2+1\right)}{x}K-2i\omega_0 H_1=\frac{2}{x}F'\ ,
\\
K''&-H''+\frac{2x}{x^2+1}(K'-H')+\omega_0^2(K+H)-\frac{2ix\omega_0 H_1}{x^2+1}-2i\omega_0 H_1'=-\frac{4}{x^2+1}F'\ ,
\\
K'&+\frac{x}{x^2+1}(K-H)-\frac{i\ell(\ell+1)}{2\omega_0\left(x^2+1\right)}H_1=-\frac{2}{x^2+1}F\ ,
\\
H_1'&+i\omega_0\left(K+H\right)=0\ ,
\\ 
K'&-H'-i\omega_0 H_1=-\frac{4}{x^2+1}F\ .
\end{align}

\begin{figure}
\centering\includegraphics[width=\hsize]{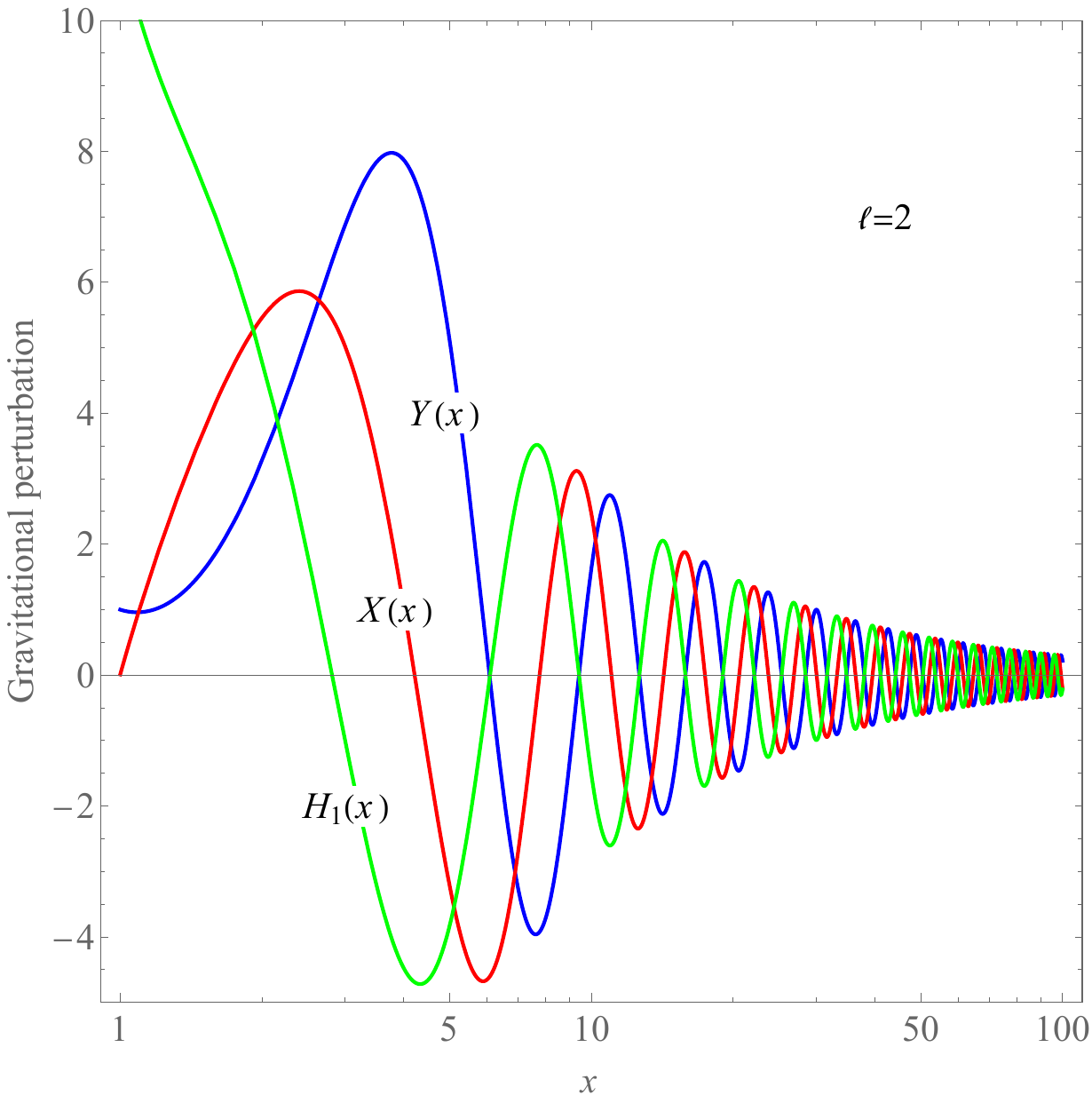}
\caption{Radial dependence of the gravitational perturbation at $\ell=2$.\label{figXY}}
\end{figure}

Hereafter introducing new radial functions $X=K+H$ and $Y=K-H$, and performing simple algebraic manipulations, one obtains

\begin{align}\label{eqY}
&Y''-\frac{2xY'}{x^2+1}+Y\left(\omega_0^2-\frac{\ell^2+\ell-2}{x^2+1}\right)=\frac{24xF}{\left(x^2+1\right)^2}\ ,
\\
&X=\frac{1}{\omega_0^2}\left[Y''+4\left(\frac{F}{x^2+1}\right)'\right]\ ,
\\
&H_1=\frac{1}{i\omega_0}\left[Y'+\frac{4}{x^2+1}F\right]\ .
\end{align}
Since the explicit expression of the function $F$ is already obtained, the task remains to solve Eq.\eqref{eqY} exclusively. Hereafter introducing a new function $Y=y\sqrt{x^2+1}$, this equation can be simplified to the Regge–Wheeler–Zerilli equation (See, e.g.\cite{Regge1957PR,Zerilli1970PRL}) including a source term in this form:
\begin{align}\label{y}
\left[\frac{d^2}{dx^2}+\omega^2-V\right]y=\frac{24xF}{(x^2+1)^{3/2}}\ ,    
\end{align}
where $V$ is the effective potential defined as
$$
V(x)=\frac{\ell(\ell+1)}{x^2}-\frac{5x^2+2}{x^2(x^2+1)^2}\ .
$$
It is difficult to find the analytical solution to equation \eqref{y}, however numerical solution of this equation can be easily obtained. As s result radial dependence of the profile function $X(x)$, $Y(x)$ and $H_1(r)$ is shown in Fig.\ref{figXY}. As one can see all profile functions oscillate with damping and disappear towards infinity. As one sees there is not any gravitational potential in the Ellis spacetime wave propagate like in a flat spacetime, in particular, if we see $r_0=0$.     

\section{Conclusion}\label{Sec:conclusions}

We have investigated Ellis spacetime by considering neutral and charged particle dynamics in the presence of the external scalar field. As well as the perturbation of the Ellis spacetime is also explored. Our main findings are summarized as follows:
\begin{itemize}
\item  
We have studied the circular motion of a massive particle in the Ellis spacetime, initially considering its potential interaction with an external scalar field for a more focused analysis. This spacetime is distinguished by two primary factors: the wormhole throat and the interaction parameter between the scalar field and the massive particle. Our study concentrated particularly on the particle's motion within the equatorial plane and aimed to ascertain how the ISCO position is influenced by these parameters. 

\item We have examined the oscillating movement of a massive particle within Ellis spacetime, focusing on understanding the equations governing its orbital and epicyclic motion. By employing perturbation techniques, we've derived linear oscillator equations describing the radial and angular displacements near the stable orbit. Exact analytical expressions for the frequencies of these oscillations along both the radial and vertical directions have been derived as well. To impose constraints on the throat of the wormhole, we've compared our theoretical findings with observational data. Our analysis showed that for low-frequency sources, the wormhole throat is no larger than $\sim 800 {\rm km}$, whereas for high-frequency sources, it resembles approximately the size of a neutron star $\sim 15 {\rm km}$. 
    
\item 
We also examined a charged particle motion around the Ellis wormhole, presuming it to be situated within a consistently uniform magnetic field. Employing the Wald method, we've presented a precise analytical solution for the azimuthal component of the vector potential of the electromagnetic field, derived from Maxwell's equations. Furthermore, we've established analytical formulas for the ISCO equation, contingent upon two primary interaction factors: the scalar field parameter $\text{g}_s$ and the magnetic parameter $\omega$, governing the interaction between charged massive particles with electromagnetic field and scalar field.
    
\item Finally, we have explored scalar and gravitational perturbations in the Ellis spacetime. We assume that both scalar gravitational waves propagate at identical frequencies and expressions for these are expanded in terms of the spherical harmonics. It is shown that the equation for the scalar profile function is totally independent from the tensor profile functions, however equations for the tensor profile functions strongly depend on the scalar profile functions in the Ellis spacetime. We have discovered that time-independent solutions for scalar and gravitational disturbances can be expressed using Legendre and associated Legendre functions, where the argument is complex. However, when considering stationary solutions within the wave zone, the exact analytical solution for scalar disturbances can be achieved, described by the confluent Heun function. It's noteworthy that the equations governing gravitational disturbances are considerably intricate, but they can be simplified to the familiar Regge–Wheeler–Zerilli equation for the tensor profile function. Finally, we present numerical solutions to the Regge–Wheeler–Zerilli equation for the radial functions.   

\end{itemize}

\section*{Acknowledgments}
This research was supported by the Grants F-FA-2021-510 from the Uzbekistan Ministry for Innovative Development.

\appendix

\section{List of radial equations}\label{App}

The list of Einstein-scalar field equations for unknown radial function  are  
\begin{widetext}
\begin{align}\label{eq1}
&K''+\frac{3r}{r^2+r_0^2}K'-\frac{\ell^2+\ell-2}{2\left(r^2+r_0^2\right)}K-\frac{r}{r^2+r_0^2}H_2'-\frac{\ell^2+\ell+2}{2\left(r^2+r_0^2\right)}H_2=-\frac{2r_0}{r^2+r_0^2}F'\ ,
\\&
K'-H_0'+\frac{2\omega^2\left(r^2+r_0^2\right)-\ell^2-\ell+2}{2r}K+\frac{\ell(\ell+1)}{2r}H_0-2i\omega H_1-\frac{H_2}{r}=\frac{2r_0}{r}F'\ ,
\\\nonumber&
\left[K''-H_0''+\frac{2rK'}{r^2+r_0^2}-\frac{rH_0'}{r^2+r_0^2}-\frac{rH_2'}{r^2+r_0^2}+\frac{4r_0F'}{r^2+r_0^2}-\frac{2ir\omega H_1}{r^2+r_0^2}-2i\omega H_1'+\omega^2(K+H_2)\right]P_\ell(\cos\theta)\\&=\frac{H_0-H_2}{r^2+r_0^2}\cot\theta\frac{d}{d\theta}P_\ell(\cos\theta)\ ,\label{qq}
\\\nonumber&
\left[K''-H_0''+\frac{2rK'}{r^2+r_0^2}-\frac{rH_0'}{r^2+r_0^2}-\frac{rH_2'}{r^2+r_0^2}+\frac{4r_0F'(r)}{r^2+r_0^2}-\frac{2ir\omega H_1}{r^2+r_0^2}-2i\omega H_1'+\omega^2(K+H_2)\right]P_\ell(\cos\theta)\\&=-\frac{H_0-H_2}{r^2+r_0^2}\frac{d^2}{d\theta^2}P_\ell(\cos\theta)\ ,\label{ff}
\\& 
K'+\frac{r}{r^2+r_0^2}K-\frac{i\ell(\ell+1)}{2\omega\left(r^2+r_0^2\right)}H_1-\frac{r}{a^2+r^2}H_2=-\frac{2r_0}{r^2+r_0^2}F\ ,
\\&
H_1'+i\omega\left(K+H_2\right)=0\ ,
\\& 
K'-H_0'+\frac{r}{r^2+r_0^2}H_0-\frac{r}{r^2+r_0^2}H_2-i\omega H_1=-\frac{4r_0}{r^2+r_0^2}F\ ,\label{KG1}
\\& 
\left[\left(r^2+r_0^2\right)F'\right]'+\left[\omega^2\left(r^2+r_0^2\right)-\ell(\ell+1)\right]F=r_0\left(K'-i\omega H_1-\frac{H_0'+H_2'}{2}\right)\ .\label{KG}
\end{align}
From Eqs. \eqref{qq} and \eqref{ff}, one can easily find that $H_0=H_2=H$ which is useful in further calculations.
\end{widetext}

\bibliographystyle{apsrev4-2}  
\bibliography{Ref}
\end{document}